\documentclass[11pt]{article}%
\usepackage{amsmath}%
\usepackage{amsfonts}%
\usepackage{amssymb}%
\usepackage{graphicx}
\providecommand{\U}[1]{\protect\rule{.1in}{.1in}}
\begin{document}

\title{Hyperfine structure and hyperfine anomaly in Pb}
\author{J. R. Persson\\Norwegian University of Science and Technology\\N-7491 Trondheim\\Norway\\jonas.persson@ntnu.no}
\maketitle

\begin{abstract}
The hyperfine structure in the $6p^{2}$-configuration in lead has been
analysed and the results is compared with calculations. The hyperfine anomaly
and improved values of the nuclear magnetic moment for four lead isotopes is
obtained, using the results from the analysis. The results open up for new
measurements of the hyperfine structure in unstable lead isotopes, in order to
extract information of the hyperfine anomaly and distribution of magnetisation
in the nucleus.

\end{abstract}

%\PACS{{31.30 Gs, 32.10 Fn}  {}

\section{Introduction}

Hyperfine structures (hfs) and isotope shifts (IS) in PbI have been studied
over the years using different techniques. The electronic ground configuration
of lead is $6p^{2}$, which gives rise to five low-lying, even-parity,
metastable states: $^{1}S_{0},^{3}P_{0,1,2},^{1}D_{2}$. The first odd-parity
state ($6p7s$ $^{3}P_{0}$) has an energy of 34960 $cm^{-1}$, which places most
transitions from the metastable states in the ultraviolet region. This has
made high-resolution laser spectroscopy difficult until the advent of
frequency doubled cw titanium-sapphire lasers. The high-lying metastable
$^{1}D_{2}$ state is accessible through transitions in the IR. The
experimental results obtained by different authors is reviewed and an analysis
based on the effective operator approach is performed in the $6p^{2}%
$configuration. The result of the analysis is used to investigate the
state-dependent hyperfine anomaly with application to unstable isotopes. The
hyperfine anomaly of four lead isotopes and improved values of the nuclear
magnetic moment has been obtained.

\bigskip

\section{Experimental hyperfine structure constants\bigskip}

\bigskip The hfs in Pb has been studied by different methods over the years,
using optical spectroscopy as well as with the Atomic Beam Magnetic Resonance
(ABMR) technique. With the advent of lasers, especially in the UV-region, more
studies have been done. In table \ref{Table1} \ an overview of the
experimental hfs constants in $^{207}Pb$ for a number of states of interest
are given. The hfs constants from high accuracy measurements have been
corrected with respect to the non-diagonal hyperfine interaction.

\begin{table}[ptbh]
\caption{Magnetic hyperfine structure constants A of the $^{207}$Pb levels, in
MHz. }%
\label{Table1}%
\begin{tabular}
[c]{|c|c|c|c|c|}\hline
Designation & A (ABMR) & A (corrected) & A (Laser) & A (Laser)\\\hline
$6s^{2}6p^{2}$ $^{3}P_{1}$ & -2390.976(2)\cite{Bengtsson} & -2390.881(2) &
-2389.4(7)\cite{Bouazza01} & -2388.2(4.5)\cite{Reeves91}\\\hline
$6s^{2}6p^{2}$ $^{3}P_{2}$ & 2602.060(1)\cite{Bengtsson} & 2602.144(1) &
2600.8(9)\cite{Bouazza01} & \\\hline
$6s^{2}6p^{2}$ $^{1}D_{2}$ & 609.818(8)\cite{Lurio 69} & 609.818(8) &  &
\\\hline
$6s^{2}6p7s$ $^{3}P_{1}$ &  &  & 8802.0(1.6)\cite{Bouazza01} &
8807.2(3.0)\cite{Anselment}\\\hline
\end{tabular}
\end{table}

There also exist studies of the hfs in unstable isotopes
\cite{Dutta91,Seliverstov09,Anselment}. A compilation of the hfs constants
obtained is given in table \ref{Table2}. These studies has mainly been
concerned with the IS, i.e. the change in nuclear charge radii, hence the use
of states without hfs.

\bigskip\begin{table}[ptbh]
\caption{Magnetic hyperfine structure constants A and B of the unstable
isotopes in lead, in MHz. }%
\label{Table2}%
\begin{tabular}
[c]{|c|c|c|c|c|c|c|}\hline
Isotope & I & A ($6p^{2}~^{1}D_{2})$ & B ($6p^{2}~^{1}D_{2})$ & A
($6p7s~^{3}P_{1})$ & B ($6p7s~^{3}P_{1})$ & Ref.\\\hline
$^{183}Pb$ & $3/2$ &  &  & -5742(25) & 70(200) & \cite{Seliverstov09}\\\hline
$^{183m}Pb$ & $13/2$ &  &  & -1423(6) & -200(400) & \cite{Seliverstov09}%
\\\hline
$^{185}Pb$ & $3/2$ &  &  & -5652(25) & -30(150) & \cite{Seliverstov09}\\\hline
$^{185m}Pb$ & $13/2$ &  &  & -1405(12) & -110(150) & \cite{Seliverstov09}%
\\\hline
$^{187}Pb$ & $3/2$ &  &  & -5580(10) & 50(200) & \cite{Seliverstov09}\\\hline
$^{187m}Pb$ & $13/2$ &  &  & -1383(5) & 60(300) & \cite{Seliverstov09}\\\hline
$^{189}Pb$ & $3/2$ &  &  & -5360(40) & -60(200) & \cite{Seliverstov09}\\\hline
$^{189m}Pb$ & $13/2$ &  &  & -1360(10) & 150(40) & \cite{Seliverstov09}%
\\\hline
$^{191}Pb$ & $13/2$ & -91.3(6) & 123(18) & -1344(1) & -15(8) & \cite{Dutta91}%
\\\hline
$^{193}Pb$ & $13/2$ & -89.6(5) & 282(14) & -1321(1) & -19(7) & \cite{Dutta91}%
\\\hline
$^{195}Pb$ & $13/2$ & -88.1(6) & 442(19) & -1294(1) & -33(9) & \cite{Dutta91}%
\\\hline
$^{197}Pb$ & $3/2$ &  &  & -5327(11) & 9(20) & \cite{Anselment}\\\hline
$^{197m}Pb$ & $13/2$ & -85.7(9) & 546(23) & -1261(7) & -59(12) &
\cite{Dutta91}\\\hline
&  &  &  & -1263(3) & -54(39) & \cite{Anselment}\\\hline
$^{199}Pb$ & $3/2$ &  &  & -5322(6) & -9(10) & \cite{Anselment}\\\hline
$^{201}Pb$ & $5/2$ &  &  & 2007.5(1.3) & 1(5) & \cite{Anselment}\\\hline
$^{202m}Pb$ & $9$ &  &  & -187.9(5) & -67(9) & \cite{Anselment}\\\hline
$^{203}Pb$ & $5/2$ &  &  & 2040.3(1.3) & -11(6) & \cite{Anselment}\\\hline
$^{205}Pb$ & $5/2$ &  &  & 2115.7(8.0) & -26(4) & \cite{Anselment}\\\hline
$^{209}Pb$ & $9/2$ &  &  & -2433(3) & 31(19) & \cite{Anselment}\\\hline
$^{211}Pb$ & $9/2$ &  &  & -2318.3(1.3) & 10(13) & \cite{Anselment}\\\hline
\end{tabular}
\end{table}

As can be seen the hfs is known in only one state for most isotopes, with the
exception of four isotopes. As we are interested in the hyperfine anomaly
these isotopes will be studied in detail.

\section{Analysis of hyperfine structure}

\subsection{\bigskip Eigenvectors}

Lead has a quite simple ground electronic configuration, but deviates from
pure LS-coupling. In order to perform an analysis of the hfs, the breakdown of
LS-coupling must be taken into account and eigenvectors have to be obtained.

The eigenvectors can be obtained by diagonalising the energy matrix of the
spin-orbital and the electrostatic interactions or by an analysis of the
experimental $g_{J}$ factors. The energy matrix has been derived by for
example Condon and Shortley \cite{Condon}. The agreement between the fitted
and experimental energy levels using this energy matrix is not particular
good. Landman and Lurio \cite{Lurio 69} included spin-spin, orbit-orbit and
spin-other-orbit interactions but this did not improve the fit.

\bigskip Instead of using the energy matrix, an analysis of the experimental
$g_{J}$ factors will probably give a better description of the system. The
experimental $g_{J}$ factors can be described as:%

\begin{equation}
g_{J}^{\exp}=\alpha^{2}g_{J}^{LS}(LS)+\beta^{2}g_{J}^{LS}(L^{\prime}S^{\prime
})\label{gj}%
\end{equation}

where $g_{J}^{LS}$ is the Lande $g_{J\text{ }}$factor for a pure LS-state
corrected for the anomalous spin of the electron, $\alpha$ and $\beta$\ are
the intermediate coupling coefficients.

The experimental $g_{J\text{ }}$ factors must be corrected for diamagnetic and
relativistic effects \cite{Judd61} \cite{Persson93}. In lead, these
corrections are of the order $5\cdot10^{-4}$, as can be seen when comparing
the experimental $g_{J\text{ }}$ factor for the $^{3}P_{1}$ state with the
corrected $g_{J\text{ }}$ factor in table \ref{Table3}. A Hartree-Fock
calculation of these diamagnetic and relativistic corrections has been done in
\cite{Persson93} and the result is presented in table \ref{Table3}. In order
to exclude coupling effects the sum of the $g_{J\text{ }}$ factor for the J=2
states are given.

\bigskip\begin{table}[ptbh]
\caption{Experimental $g_{J}$ factors compared with values corrected for
relativistic and diamagnetic effects.}%
\label{Table3}%
\begin{tabular}
[c]{|c|c|c|c|c|}\hline
State & Lande value & Corrections\cite{Persson93} & Calculated &
Experimental\cite{Bengtsson}\cite{Lurio 69}\\\hline
$^{3}P_{1}$ & 1.5011596 & -0.0001060 & 1.5010536 & 1.500755(10)\\\hline
$^{1}D_{2}+^{3}P_{2}$ & 2.5011596 & -0.0002783 & 2.5008813 &
2.50148(11)\\\hline
\end{tabular}
\end{table}

\bigskip

The calculated corrections were not as large as expected, why configuration
interaction effects should be important. It has been shown by Gil and Heldt
\cite{Gil} that there exists a configuration mixing between the $6p^{2}$ and
$6p7p$ configurations, by including configuration interactions in the energy
matrix analysis. Even though their fit suffer from the same problems as in the
ordinary matrix analysis, a calculation of the $g_{J\text{ }}$ factors using
their eigenvectors and including diamagnetic and relativistic corrections gave
an excellent agreement in comparison with experimental data \cite{Persson93}.

In this case we exclude the configuration interaction when analysing the
$g_{J\text{ }}$ factors, as a precaution, in order to obtain accurate
eigenvectors, the estimated errors of the relativistic and diamagnetic
corrections were enlarged.

All obtained eigenvectors are given in table \ref{Table 4}. In case A the
eigenvectors are obtained by analysing the energy levels according to the
energy matrix of Condon and Shortley \cite{Condon}, in case B eigenvectors are
derived by Landman and Lurio \cite{Lurio 69} and in case C the eigenvectors
are obtained by analysing the experimental $g_{J\text{ }}$ factors.

\begin{table}[ptbh]
\caption{Obtained eigenvectors}%
\label{Table 4}%
\begin{tabular}
[c]{|c|c|c|}\hline
& $\alpha$ & $\beta$\\\hline
Case A & 0.765717 & 0.643178\\\hline
Case B & 0.7636 & 0.6457\\\hline
Case C & 0.740780(23) & 0.671748(158)\\\hline
\end{tabular}
\end{table}

\bigskip

\subsection{\bigskip Hyperfine interaction}

The analysis of the hyperfine interaction is based on an effective hyperfine
hamiltonian, which for the magnetic dipole interaction is written as
\cite{lindgrenrosen}:%

\begin{equation}
H_{hfs,eff}^{1}=2\frac{\mu_{0}}{4\pi}\mu_{B}\sum\limits_{i=1}^{N}\left[
\mathbf{l}_{i}\cdot\left\langle r^{-3}\right\rangle ^{01}-\sqrt{10}\frac
{g_{s}}{2}\left(  \mathbf{sC}^{2}\right)  _{i}^{1}\cdot\left\langle
r^{-3}\right\rangle ^{12}+\frac{g_{s}}{2}\mathbf{s}_{i}\cdot\left\langle
r^{-3}\right\rangle ^{10}\right]  \bullet\mathbf{M}^{1}\label{eff hfs}%
\end{equation}

\bigskip By determination of the angular parts, using the eigenvectors, the
magnetic dipole interaction constants "A" can be expressed as a linear
combination of the orbital (01), spin-dipole (12), and contact (10) effective
radial parameters $(a^{ij})$. \bigskip%
\begin{equation}
A=k^{01}a^{01}+k^{12}a^{12}+k^{10}a^{10}%
\end{equation}

The numbers in the parentheses correspond to the rank of the spherical tensor
operators in the spin and orbital spaces. In this way can the effective radial
parameters for the different eigenvectors be fitted to the corrected A factors.

\bigskip

The obtained effective radial parameters are presented in table \ref{table 5}.
The errors in the effective radial parameters are mainly due to uncertainty of
the eigenvectors, since the errors in the energy fit is quite large and hard
to obtain, these errors are expected to be on the order of 10\%. In the
analysis of the experimental $g_{J}$ factors, the errors are possible to
obtain from the fit.

\bigskip\begin{table}[ptbh]
\caption{Values of the effective radial parameters in MHz}%
\label{table 5}%
\begin{tabular}
[c]{|c|c|c|c|}\hline
& $a^{01}$ & $a^{12}$ & $a^{10}$\\\hline
Case A & 2365.68 & 5395.26 & -1752.18\\\hline
Case B & 2377.74 & 5375.16 & -1784.34\\\hline
Case C & 2518.80(88) & 5134.10(88) & -2158.01(12)\\\hline
\end{tabular}
\end{table}

\bigskip

\bigskip The effective radial parameters, proportional to the nuclear moment
and the effective $\left\langle r^{-3}\right\rangle $ values can be expressed
as \cite{lindgrenrosen}:

\bigskip%
\begin{equation}
a^{ij}=2\frac{\mu_{0}}{4\pi}\mu_{B}\frac{\mu_{I}}{I}\left\langle
r^{-3}\right\rangle ^{ij}\label{r-3}%
\end{equation}

\bigskip Since the nuclear magnetic dipole moment has been determined
independently ($\mu_{I}=+0.592583(9)$ n.m.), it is possible to derive the
effective $\left\langle r^{-3}\right\rangle $ values. These semi-empirical
values are presented in table \ref{table 6}\ together with calculated
$\left\langle r^{-3}\right\rangle $ values using the Hartree-Fock (HF) and
Optimized Hartree-Fock-Slater (OHFS) methods by Lindgren and Rosen
\cite{lindgrenrosen}.

\bigskip

\begin{table}[ptbh]
\caption{Experimental and calculated hyperfine integrals ( in units of
$a_{0}^{-3}$).}%
\label{table 6}%
\begin{tabular}
[c]{|c|c|c|c|}\hline
& $\left\langle r^{-3}\right\rangle ^{01}$ & $\left\langle r^{-3}\right\rangle
^{12}$ & $\left\langle r^{-3}\right\rangle ^{10}$\\\hline
HF\cite{lindgrenrosen} & 22.302 & 44.390 & -7.337\\\hline
OHFS\cite{lindgrenrosen} & 22.898 & 48.076 & -8.519\\\hline
Exp case A & 20.921 & 47.713 & -15.495\\\hline
Exp case B & 21.028 & 47.535 & -15.780\\\hline
Exp case C & 22.2750(84) & 45.4034(86) & -19.0844(15)\\\hline
\end{tabular}
\end{table}

\bigskip The calculated relativistic values of $\left\langle r^{-3}%
\right\rangle ^{01}$ differ from the experimental value (case C) by 1.4\% for
the OHFS and 1.3\%, for the HF method, while the corresponding difference
between the calculated and experimental values of $\left\langle r^{-3}%
\right\rangle ^{12}$ is 4.4\% and 3.7\%, respectively. The large difference
between the experimental and calculated values of $\left\langle r^{-3}%
\right\rangle ^{10}$ are mainly due to spin polarisation. Bouazza et al.
\cite{Bouazza01} estimated the fraction of the spin polarisation to be
50.62\%, as shown the isoelectronic Bi II \cite{Bouazza88}, yielding a value
of $\left\langle r^{-3}\right\rangle ^{10}=-16.06a_{0}^{-3}$, in reasonable
agreement with the experimental values.

\bigskip

\section{Hyperfine anomaly}

\bigskip In addition to the hyperfine interaction and nuclear magnetic dipole moment is it possible to obtain information on the distribution
of magnetisation in the nucleus through the so called Bohr-Weisskopf effect
(BW-effect) \cite{bohrweisskopf,Fujitaarima,Buttgenbach}. The first to consider the influence of the
finite size of the nucleus on the hyperfine structure was 
Bohr and Weisskopf \cite{bohrweisskopf}. They calculated the hyperfine
interaction of s$_{1/2}$ and p$_{1/2}$ electrons for an extended
nucleus, and showed that the magnetic dipole hyperfine interaction constant
($A$) for an extended nucleus is generally smaller than for a
point nucleus. The effect on the hyperfine interaction from the extended charge distribution of the nucleus gives rise to
the so-called Breit-Rosenthal effect
(BR-effect)\cite{Breit,Crawford,Pallas,Rosenberg}. In this case, as in most but not all
cases, the differential BR-effect is negligible when two isotopes are
compared. Inclusion of the BR-effect will not have any effect on the
results, since the BW- and BR-effects show the same behaviour. The BR-effect
is therefore neglected in the following discussion. 
Isotopic variations of
magnetic moments became larger than those in the point dipole interaction since
there are different contributions to the hfs from the orbital and spin parts
of the magnetisation in the case of extended nuclei. 
The fractional difference
between the point nucleus hfi constant ($A_{point}$) and the constant obtained
for the extended nuclear magnetisation is commonly referred to as the
Bohr-Weisskopf (BW) effect \cite{Buttgenbach}.The hfs constant $A$ can
therefore be written as%

\begin{equation}
A=A_{point}\left(  1+\epsilon_{BW}\right) \label{eq1}%
\end{equation}

where $\epsilon_{BW}$ is the BW-effect, and $A_{point}$ is the $A$ constant
for a point nucleus. The BW-effect is dependent on both nuclear and atomic
properties, i.e. the electron density within the nucleus. The nuclear part,
i.e. the distribution of nuclear magnetisation, can be calculated using
different nuclear models \cite{Fujitaarima,Buttgenbach}. Since electronic
wavefunctions cannot be calculated with sufficient high accuracy in complex atoms, as
they can be in hydrogen-like ions and muonic atoms, it is not possible
to determine\ $\epsilon_{BW}$ directly in atoms. However, it is possible to determine
the difference of the BW-effect in two isotopes, the so-called (differential)
hyperfine anomaly (hfa). Comparing the ratio of the measured hfs constants
for two isotopes with the independently measured ratio of the nuclear magnetic
dipole moments to extract the hfa,$^{1}\Delta^{2}$, for the isotopes 1 and 2,
and a given atomic state, gives:%

\begin{equation}
1+{^{1}\Delta^{2}}={{\frac{A^{(1)}}{A^{(2)}}}{\frac{{\mu_{I}^{(2)}/I^{(2)}}%
}{{\mu_{I}^{(1)}/I^{(1)}}}}\approx1+{\epsilon_{BW}^{(1)}}-{\epsilon_{BW}%
^{(2)}}}\label{eq2}%
\end{equation}

where $\mu_{I}$ is the nuclear magnetic dipole moment, and I the nuclear spin.
In the case of electrons with a total angular momentum j$>$1/2 the anomalies may be
disregarded as the corresponding wavefunctions vanish at the nucleus. The hfa
can show a dependence of the atomic state, a state dependent hfa, where the
values for different states can vary significantly. The reason for the state dependence is that the
hyperfine interaction consists of three parts \cite{Ref4,Ref5}, orbital,
spin-orbit and contact (spin) interaction, where only the contact interaction
contributes to the hfa. Since the contribution of the different interactions differ between different atomic states, and it is only the spin interaction giving rise to the hfa, a state dependent hfa is the result.  It is therefore suitable to rewrite the dipole hyperfine
interaction constant as%

\begin{equation}
A=A_{nc}+A_{c}\label{eq3}%
\end{equation}

where $A_{c}$ is the contribution due to the contact interaction of s (and
p$_{1/2}$) electrons and $A_{nc}$ is the contribution due to non-contact
interactions. The experimental hfa, which is defined with the total magnetic
dipole hyperfine constant $A$, should then be rewritten to obtain the relative
contact contribution to the hfa:%

\begin{equation}
{^{1}\Delta_{exp}^{2}}={^{1}\Delta_{c}^{2}}{\frac{A_{c}}{A}}\label{eq4}%
\end{equation}

where ${^{1}\Delta_{c}^{2}}$ is the hfa due to the contact interaction, that
is, for an s- or p$_{1/2}$-electron.

From the discussion, one might come to the conclusion that one needs independent
measurements of the nuclear magnetic moments and the $A$-constants in order to
obtain the hfa, however, this is not true. As has been shown by Persson
\cite{persson}, it is possible to extract the anomaly solely from the
$A$-constants of two different atomic states, provided the ratio $\left(
\frac{A_{c}}{A}\right)  $differs  for the different states.
Comparing the A-constant ratio, for two isotopes, in two atomic states, gives:%

\begin{equation}
{\frac{{A_{B}^{(1)}/A_{B}^{(2)}}}{{A_{C}^{(1)}/A_{C}^{(2)}}}}\approx
{1+{^{1}\Delta_{c}^{2}}({{\frac{A_{c}^{B}}{A^{B}}}-{\frac{A_{c}^{C}}{{A^{C}}}%
})}}\label{eq5}%
\end{equation}

Where B and C denotes different atomic states and 1 and 2 denotes different
isotopes. The ratio between the two $A$-constant ratios for the isotopes will
therefore only depend on the difference in the contact contributions of the two atomic
states and the hfa. It should also be noted that the ratio $\left(  \frac{A_{s}}%
{A}\right)  $ is isotope independent. Once determined for one isotopic pair,
the ratio can be used for all pairs, which is very useful in the study of hfa in
radioactive isotopes. It is possible to determine the ratio in two different ways;
either by an analysis of the hyperfine interaction or by using a known
hfa as a calibration. It should be noted that the atomic states used
must differ significantly in the ratio$\left(  \frac{A_{s}}{A}\right)  $, as a
small difference will lead to an increased sensitivity to errors.

Since the hfa is normally very small (1\% or less) it is often necessary to have
high accuracy for the $A$-constants , preferably better than 10$^{-4}$ \cite{Buttgenbach}%
. In stable isotopes there is no major problem to measure the
nuclear magnetic moment with sufficient accuracy using NMR or ABMR, while for unstable isotopes it is
more difficult. In most cases there does not exist any high precision
measurements of the nuclear magnetic moment, in most cases the nuclear magnetic moment is deduced from the hfs while neglecting the effect of hfa. However, there might exist
measurements of two $A$-constants, if the nuclear charge radius of the
unstable isotopes has been measured by means of laser spectroscopy. In order
to obtain the hfa one therefore needs to measure the $A$-constants with an accuracy
better than 10$^{-4}$, something that can be done by laser spectroscopy provided the
$A$-constant is larger than about 1000 MHz, as being the case in Pb.

\bigskip

\section{Hyperfine anomaly in unstable isotopes}

\bigskip From table \ref{Table2} we see that the A constants are known for two
states in four unstable isotopes, $^{191}Pb$,$^{193}Pb$,$^{195}Pb$ and
$^{197m}Pb$. The complication is that one state has a small A constant and the
other belongs to the 6p7s configuration. Still, it is possible to obtain a
state dependent hyperfine anomaly using:%

\begin{equation}
{\frac{{A_{B}^{(1)}/A_{B}^{(2)}}}{{A_{C}^{(1)}/A_{C}^{(2)}}}}\approx
{1+{^{1}\Delta_{\exp}^{2}}}%
\end{equation}

\bigskip with the A constants from $^{207}$Pb, as reference nucleus (1). The
state dependent hyperfine anomalies obtained are given in table \ref{Table 7}.

\bigskip\begin{table}[ptbh]
\caption{State dependent hyperfine anomaly. }%
\label{Table 7}%
\begin{tabular}
[c]{|c|c|}\hline
Isotope & $^{207}\Delta_{\exp}^{A}(\%)$\\\hline
$^{191}Pb$ & -1.94(68)\\\hline
$^{193}Pb$ & -2.10(58)\\\hline
$^{195}Pb$ & -1.73(70)\\\hline
$^{197m}Pb$ & -1.90(123)\\\hline
\end{tabular}
\end{table}

\bigskip Note that the hfa contains contributions from both states involved.
This makes the contact contribution of the hyperfine interaction is quite
complicated with both $s$ and $p_{1/2}$ electrons. However it is possible to
examine the hyperfine interaction in the $6p7s$ $^{3}P_{1}$ further. Bouazza
et al \cite{Bouazza01} gives the eigenvector components for this state, and by
assuming that the effective hyperfine interaction parametes for the p
electrons are the same in the $6p^{2}$ and $6p7s$ configurations, we can
deduce a value of the s electron effective hyperfine interaction parameter.

Using the eigenvector we find that the A constant for the $6p7s$ $^{3}P_{1}$
can be expressed in effective hyperfine interaction parameters as:

\bigskip%
\begin{subequations}
\begin{align}
{A(}6p7s^{3}P_{1})  & =0.63815a_{p}^{01}+0.67804a_{p}^{12}-0.13526a_{p}%
^{10}+0.49712a_{s}^{10}\\
& =8802.0(1.6)MHz
\end{align}
\newline

Using the effective hyperfine interaction parameters for the p electrons in
table \ref{table 5}, gives the s electron parameter $a_{s}^{10}=6884$ $MHz$ ,
which is a reasonable value. It is now possible to calculate the contact
contribution in equation \ref{eq5}, both for the $6p7s$ $^{3}P_{1}$ and
$6p^{2}$ $^{1}D_{2}$ states.

\bigskip%
\end{subequations}
\begin{align}
\frac{{A}_{{c}}}{A}{(}6p7s^{\text{ }3}P_{1})  & =0.422\\
\frac{{A}_{{c}}}{A}{(}6p^{2\text{ }1}D_{2})  & =-0.708
\end{align}

Using this it is possible to obtain the state independent contact anomaly. It
must be noted that the contact anomaly consists of both $s$- and $p_{1/2}%
$-electron parts. If we assume that the contribution to the hyperfine anomaly
is the same for s- and p- electrons we must correct the obtained state
dependent hyperfine anomalies by a factor 1.13 ($\frac{{A}_{{c}}}{A}%
{(}6p7s^{\text{ }3}P_{1})-\frac{{A}_{{c}}}{A}{(}6p^{2\text{ }1}D_{2})$),
giving the state independent hyperfine anomaly in table \ref{table 7a}.

\bigskip\begin{table}[ptbh]
\caption{State independent hyperfine anomaly. }%
\label{table 7a}%
\begin{tabular}
[c]{|c|c|}\hline
Isotope & $^{207}\Delta_{c}^{A}(\%)$\\\hline
$^{191}Pb$ & -1.72(68)\\\hline
$^{193}Pb$ & -1.86(58)\\\hline
$^{195}Pb$ & -1.53(70)\\\hline
$^{197m}Pb$ & -1.68(123)\\\hline
\end{tabular}
\end{table}

In order to check if the result is reasonable we can use the obtained
hyperfine anomaly to calculate the nuclear magnetic dipole moment of the four
isotopes using the measured A constants:%

\begin{equation}
{{\frac{A^{(1)}}{A^{(2)}}}{\frac{{\mu_{I}^{(2)}/I^{(2)}}}{{\mu_{I}%
^{(1)}/I^{(1)}}}=}1+{^{1}\Delta_{c}^{2}}{\frac{A_{c}}{A}}}%
\end{equation}

rearranging gives%

\begin{equation}
{\mu_{I}^{(2)}{=}}\frac{A^{(2)}}{A^{(1)}}\frac{{I^{(2)}}}{{I^{(1)}}}{\mu
_{I}^{(1)}}\left(  {1+{^{1}\Delta_{c}^{2}}{\frac{A_{c}}{A}}}\right) \label{15}%
\end{equation}

Using this we can calculate the nuclear magnetic dipole moment using both
atomic states, the results are given in table \ref{Table 8}.

\begin{table}[ptbh]
\caption{Nuclear magnetic dipole moments from \cite{Dutta91} and derived
correcting for the hyperfine anomaly. The errors are only from experimental
uncertainty. }%
\label{Table 8}%
\begin{tabular}
[c]{|c|c|c|c|c|}\hline
Isotope & $\mu_{I}(^{1}D_{2})$ & $\mu_{I}(^{1}D_{2})\cite{Dutta91}$ & $\mu
_{I}(^{3}P_{1})$ & $\mu_{I}(^{3}P_{1})\cite{Dutta91}$\\\hline
$^{191}Pb$ & -1.167(15) & -1.155(15) & -1.168(8) & -1.176(8)\\\hline
$^{193}Pb$ & -1.147(14) & -1.133(14) & -1.147(8) & -1.156(8)\\\hline
$^{195}Pb$ & -1.125(15) & -1.114(15) & -1.125(8) & -1.132(8)\\\hline
$^{197m}Pb$ & -1.095(19) & -1.084(19) & -1.094(14) & -1.103(14)\\\hline
\end{tabular}
\end{table}

The agreement between the different states is much better for the corrected
values, giving a better value for the nuclear magnetic moments.

\subsection{\bigskip The empirical Moskowitz-Lombardi formula.}

The empirical Moskowitz-Lombardi (ML) formula was established in 1973 as a
rule for the s-electron BW-effect in mercury isotopes\cite{ML}.%

\begin{equation}
\epsilon_{BW}{=}\frac{{\alpha}}{\mu_{I}},\alpha=\pm1.0\cdot10^{-2}\mu
_{N},I=l\pm\frac{1}{2}%
\end{equation}

where $l$ is the orbital momentum for the odd neutron. It turned out that the
empirical rule provided a better agreement with experimental hfa than the
theoretical calculations performed by Fujita and Arima \cite{Fujitaarima}
using microscopic theory. The rule can be qualitatively explained by the
microscopic theory used by Fujita and Arima \cite{Fujitaarima}, where the
parameter $\alpha$ is more state independent than given by the theory. Further
investigations gave an analogous expression for the odd-proton nuclei
$^{191,193}$Ir , $^{197,199}$Au and $^{203,205}$Tl, but also for the
doubly-odd $^{196,198}$Au nuclei. The results indicate that the spin operators
$g_{s}^{\left(  i\right)  }\Sigma_{i}^{\left(  1\right)  }$are state
independent for these nuclei. It is worth noting that all nuclei discussed lie
close to the doubly closed shell nucleus $^{208}$Pb, where one would expect
the single particle model to provide a good description of the nucleus. In our
case we would expect a fair agreement with the ML formula. However we have to
keep in mind that the contact contribution is not a pure s anomaly, thus there
will be an unknown numerical factor. We also have to take into account that
the nuclear configuration of the unstable isotopes ( $i_{13/2},I=l+\frac{1}%
{2}$) and the reference nucleus ($^{207}Pb,p_{1/2},I=l-\frac{1}{2}$) are
different. The ML formula for calculating the hyperfine anomaly will therefore be:%

\begin{subequations}
\begin{align}
^{207}\Delta_{c}^{A}  & =\epsilon_{207}-\epsilon_{A}=\left(  \frac
{-1.0\cdot10^{-2}\mu_{N}}{\mu_{I}^{207}}-\frac{1.0\cdot10^{-2}\mu_{N}}{\mu
_{I}^{A}}\right) \\
& =-1.0\cdot10^{-2}\mu_{N}\left(  \frac{1}{\mu_{I}^{207}}+\frac{1}{\mu_{I}%
^{A}}\right)
\end{align}

\bigskip The calculated hyperfine anomalies are given in table \ref{Table9}%
\ with the experimental hyperfine anomalies.

\bigskip\begin{table}[ptbh]
\caption{Hyperfine anomalies calculated with the ML\ formula and experimental
hyperfine anomalies}%
\label{Table9}%
\begin{tabular}
[c]{|c|c|c|}\hline
Isotope & $^{207}\Delta_{ML}^{A}$ & $^{207}\Delta_{c}^{A}$\\\hline
$^{191}Pb$ & -0.938 & -1.72(68)\\\hline
$^{193}Pb$ & -0.920 & -1.86(58)\\\hline
$^{195}Pb$ & -0.902 & -1.53(70)\\\hline
$^{197m}Pb$ & -0.874 & -1.68(123)\\\hline
\end{tabular}
\end{table}

From the results it seems that the ML formula still holds. The value of
$\alpha$ seems to be too small, but one have to keep in mind that the ML
formula uses s-electrons, the contact anomaly used in this case contains both
s- and p-electron contributions, which must be evaluated further in order to
make a more quantitative comparison.

\section{Discussion}

\bigskip The hyperfine structure of $^{207}$Pb has been analysed and the
analysis has been used as the basis for determining the hyperfine anomaly in
four unstable isotopes, $^{191}Pb$,$^{193}Pb$,$^{195}Pb$ and $^{197m}Pb$,
using the method of Persson \cite{persson}. The derived hyperfine anomaly has
then been used to obtain better values of the nuclear magnetic moment for the
unstable isotopes. There exists measurements in other unstable isotopes, table
\ref{Table2}, but only in one state that exhibits hyperfine structure, why it
is not possible to derive the hyperfine anomaly in these isotopes. It would be
possible if another atomic state is measured in these isotopes, preferably the
$6p^{2}$ $^{1}D_{2}$. It is also possible to make the new measurements in the
$6p^{2}$ $^{3}P_{1}$ or $^{3}P_{2}$. The optimum would be to make measurements
in all possible states in the $6p^{2}$ configuration, thus giving in total
four atomic states that enable a cross-check of the results. Due to
constrains in the population of atomic states at accelerators, only the lower
lying states would be feasible, thus excluding the relatively high energy
state $6p^{2}$ $^{1}D_{2}$. The remaining states offer another complication,
as the contact contribution to the hyperfine structure for the $6p^{2}$
$^{3}P_{1}$ state ($\frac{A_{s}}{A}=0.452$) is close to the contribution of
the $6p7s$ $^{3}P_{1}$ state ($\frac{A_{s}}{A}=0.422$), which is not suitable
for an analysis of the hyperfine anomaly \cite{persson}. The contact
contribution to the hyperfine structure for the $6p^{2}$ $^{3}P_{2}$ state
($\frac{A_{s}}{A}=-0.247$), is suitable for analysis with all other states. An
experiment where the hyperfine structure of the $6p^{2}$ $^{3}P_{2}$ state in
unstable isotopes of lead are measured would give both better values of the
nuclear magnetic moments as well as values of the hyperfine anomaly.

The empirical Moskowitz-Lombardi formula seems to be valid, from the results
here. The mix of s- and p-electron contributions makes it difficult to make
quantitative comparison, however, with more experimental data this should be possible.

\bigskip

\end{subequations}

\end{document}